\documentclass{article}

\usepackage{arxiv}

\usepackage[utf8]{inputenc} % allow utf-8 input
\usepackage[T1]{fontenc}    % use 8-bit T1 fonts
\usepackage{hyperref}       % hyperlinks
\usepackage{url}            % simple URL typesetting
\usepackage{booktabs}       % professional-quality tables
\usepackage{amsfonts}       % blackboard math symbols
\usepackage{nicefrac}       % compact symbols for 1/2, etc.
\usepackage{microtype}      % microtypography
\usepackage{lipsum}

\usepackage{morefloats}
\usepackage{color}
\usepackage{multirow}
\usepackage{latexsym}
\usepackage{amsmath,amssymb,amsthm,graphicx}
\usepackage{dsfont}
\usepackage{enumitem}
\usepackage{hyperref}
\usepackage{arydshln}
\usepackage{lscape}
\usepackage{adjustbox}
\DeclareMathAlphabet{\mathscr}{OT1}{pzc}{m}{it}

\newtheoremstyle{def}% name
{9pt}%      Space above, empty = `usual value'
{9pt}%      Space below
{}% Body font
{}%         Indent amount (empty = no indent, \parindent = para indent)
{\bfseries}% Thm head font
{.}%        Punctuation after thm head
{ }% Space after thm head: \newline = linebreak
{}%         Thm head spec
\theoremstyle{def}

\renewcommand{\footnoterule}{%
	\kern -3.5pt
	\hrule width \textwidth height 1pt
	\kern 3.5pt
}

\makeatletter
\def\blfootnote{\xdef\@thefnmark{}\@footnotetext}
\makeatother

\title{A proposed simulation technique for population stability testing in credit risk scorecards}

\author{J. du Pisanie\\
School of Mathematical and Statistical Sciences,\\ North-West University,\\ South Africa. \\
\href{mailto:dupisanie@gmail.com}{dupisanie@gmail.com}\\
\And J. S. Allison\\
School of Mathematical and Statistical Sciences,\\ North-West University,\\ South Africa.\\
\href{mailto:james.allison@nwu.ac.za}{james.allison@nwu.ac.za}
\And I. J. H. Visagie\\
School of Mathematical and Statistical Sciences,\\ North-West University,\\ South Africa.\\
\href{mailto:jaco.visagie@nwu.ac.za}{jaco.visagie@nwu.ac.za}\\
}

\begin{document}

\date{\today}
\maketitle

\begin{abstract}

Credit risk scorecards are logistic regression models, fitted to large and complex data sets, employed by the financial industry to model the probability of default of a potential customer. In order to ensure that a scorecard remains a representative model of the population one tests the hypothesis of population stability; specifying that the distribution of clients' attributes remains constant over time. Simulating realistic data sets for this purpose is nontrivial as these data sets are multivariate and contain intricate dependencies. The simulation of these data sets are of practical interest for both practitioners and for researchers; practitioners may wish to consider the effect that a specified change in the properties of the data has on the scorecard and its usefulness from a business perspective, while researchers may wish to test a newly developed technique in credit scoring.

We propose a simulation technique based on the specification of bad ratios, this is explained below. Practitioners can generally not be expected to provide realistic parameter values for a scorecard; these models are simply too complex and contain too many parameters to make such a specification viable. However, practitioners can often confidently specify the bad ratio associated with two different levels of a specific attribute. That is, practitioners are often comfortable with making statements such as ``on average a new customer is 1.5 times as likely to default as an existing customer with similar attributes''. We propose a method which can be used to obtain parameter values for a scorecard based on specified bad ratios. The proposed technique is demonstrated using a realistic example and we show that the simulated data sets adhere closely to the specified bad ratios. The paper provides a link to a github project in which the R code used in order to generate the results shown can be found.

%As a point of independent interest, we note that the proposed method is not only useful in credit scoring, but also has applications in a wide range of other models. We outline some of these applications as directions for further research.
\vspace{0.5cm}



\end{abstract}

\section{Introduction and motivation}
\label{Intro}

%this hypothesis specifies that the current proportions of the population in the various risk buckets are the same as was the case at the point in time at which the scorecard was developed.

Credit scoring is an important technique used in many financial institutions in order to model the probability of default, or some other event of interest, of a potential client. For example, a bank typically has access to data sets containing information pertinent to credit risk which may be used in order to assess the credit worthiness of potential clients. The characteristics or covariates recorded in such a data set are referred to as attributes throughout; these include information such as income, the total amount of outstanding debt held and the number of recent credit enquiries. A bank may use logistic regression to model an applicant's probability of default as a function of their recorded attributes; these logistic regression models are referred to as credit risk scorecards. In addition to informing the decision as to whether or not a potential borrower is provided with credit, the scorecard is typically used to determine the quoted interest rate. For a detailed treatment of scorecards, see \cite{SIDDIQI2006} as well as \cite{SIDDIQI2016}.

The development of credit risk scorecards are expensive and time consuming. As a result, once properly trained and validated, a bank may wish to keep a scorecard in use for an extended period, provided that the model continues to be a realistic representation of the attributes of the applicants in the population. One way to determine whether or not a scorecard remains a representative model is to test the hypothesis of population stability. This hypothesis states that the distribution of the attributes remains unchanged over time (i.e., that the distribution of the attributes at present is the same as the distribution observed when the scorecard was developed). When the distribution of the attributes change, it provides the business with an early indication that the scorecard may no longer be a useful model.

In order to perform scenario testing in the context of population stability, the ability to simulate realistic data sets would certainly be helpful; this paper proposes a simple technique for the simulation of such data sets. This allows practitioners to consider scenarios with predefined deviations from specified distributions, they may then gauge the effect that changes in the distribution of one or more attributes has on the predictions of the model. The business may also wish to test the effects of a certain strategy in advance. For example, if a bank markets aggressively to younger people, they may wish to test the effect of a shift in the distribution of the age of their clients.

The concept of population stability can be further illustrated by means of a simple example. Consider a model that predicts whether someone is wealthy based on a single attribute; the value of the property owned. If this attribute exceeds a specified value, the model predicts that a person is wealthy. Due to house price inflation, the overall prices of houses rise over time. Thus, after a substantial amount of time has passed, the data can no longer be interpreted in the same way as before and the hypothesis of population stability is rejected, meaning that a new model (or perhaps just a new cut off point) is required.

Population stability metrics measure the magnitude of the change in the distribution of the attributes over time.
%This is done to identify substantial changes in the distribution of the attributes so as to have an indication that the model might no longer provide accurate predictions of the probability of default.
%In the above example it can be expected that the proportion of people owning a house worth more than the indicated value constantly increases. This could be an early indication that the model output is no longer reliable.
A number of techniques have been described in the literature whereby population stability may be tested; see \cite{TAPLINHUNT2019}, \cite{YN2019} as well as \cite{DUPISANIEVISAGIE2020}. For practical implementations of techniques for credit risk scorecards, see \cite{Fan2020} in the statistical software R as well as
\cite{Pru2010} in Statistical Analysis Software (SAS). The mentioned papers typically provide one or more numerical examples illustrating the use of the proposed techniques. The data sets upon which these techniques are used are typically protected by regulations, meaning that including examples based on observed data is problematic. As a result, authors often use simulated data. However, the settings wherein these examples are to be found are often oversimplified, stylised and not entirely realistic. This can, at least in part, be ascribed to the difficulties associated with the simulation of realistic data sets. These difficulties arise as a result of the complexity of the nature of the relationship between the attributes and the response.
%In this paper, we propose a technique which may be used in order to simulate realistic data sets in the context of credit risk scorecards.

%A model often encountered in practice is a scorecard, especially in a credit risk environment. A scorecard is a logistic model used to predict the probability of a binary outcome, for example the event of eventual default. Although the techniques described in the remainder of the paper can be applied in a quite general setting, we illustrate their use with scorecards.

The data sets typically used for scorecard development have a number of features in common. They are usually relatively large; typical ranges for the number of observations range from one thousand observations to one hundred thousand, while a sample size of one million observations is not unheard of. The data used are multivariate; the number of attributes used varies according to the type of scorecard, what the scorecard will be used for and other factors, but scorecards based on five to fifteen attributes are common. The inclusion of attributes in a scorecard depends on the predictive power of the attribute as well as more practical considerations. These can include the ability to obtain the required data in future (for example, changing legislation may, in future, prohibit the inclusion of certain attributes such as gender into the model) as well as the stability of the attribute over the expected lifetime of the scorecard. Care is usually taken to include only attributes with a low level of association with each other so as to avoid the problems associated with multicolinearity.

%\textcolor{red}{It is common practice, see [Siddiqi] for attributes to be included ``which are independent of each other''. [Have a look at eqactly what Siddiqi has to say on the subject.] Scorecards are typically built and used under the assumption that the attributes are independent [no interaction term is used].}

%\textcolor{red}{Okay... so this took me a few weeks to formulate... When testing population stability, we are testing whether or not the assumed characteristics underlying the model hold for the population to which the model will be applied. This is not exactly the same as testing whether or not the characteristics of the current population match those of the population on which the model was built. The difference is subtle, but  has far reaching implications. For example, while the attributes used in the model are chosen so as to be approximately uncorrelated, these attributes will not be independent. However, the assumptions of the model includes independence. Therefore, under the null hypothesis, we should simulate attributes independent from each other.}

This paper proposes a simple simulation technique which may be used to construct realistic data sets containing attributes as well as outcomes in the credit risk setting. The constructed data sets can be used to empirically investigate the effects of changes in the distribution of the attributes as well as changes in the relationship between the attributes and the event of interest. It should be noted at the outset that the proposed technique is not restricted to the context of credit scoring, or even to the case of logistic regression, but rather has a large number of other modelling applications. However, we restrict our attention to this important special case for the remainder of the paper.
%Some other applications are outlined in the concluding section together with directions for further research.

The premise of the proposed simulation technique outlines as follows. When building a scorecard, practitioners cannot be relied upon to specify the parameters of the model which will ultimately be used to model the data. The large number of parameters in the model coupled with the complex relationships between these parameters conspire to make this task near impossible. However, practitioners can be called on to have intuition regarding the bad ratios associated with different states of an attribute; i.e., practitioners are often comfortable making statements like ``on average new customer is 1.5 times as likely to default as existing customers with similar attributes''. This paper proposes a technique which can be used to choose parameter values which mimic these specified bad ratios. The inputs required for the proposed technique are the overall bad rate, the specified bad ratios and the marginal distributions of the attributes. It should be noted that not all users of scorecards are trained in statistics, meaning that the simple nature of the proposed simulation technique (i.e., specifying bad ratios and choosing parameters accordingly) is advantageous.

The remainder of the paper is structured as follows. Section 2 shows several examples of settings in which logistic regression is used in order to model the likelihood of an outcome based on attributes. Here we demonstrate the need for the proposed simulation procedure. A realistic setting is specified in this section which is used throughout the paper. Section 3 proposes a method that may be used to translate specified bad ratios into model parameters emulating these bad ratios using simulation, followed by parameter estimation. We discuss the numerical results obtained using the proposed simulation technique in Section 4. Section 5 provides some conclusions as well as directions for future research.

%We are interested in simulating attributes and outcomes from a population with specified characteristics. Below, we propose a technique that chooses parameter values that makes the specified bad rations work good.

%When using logistic regression, we assume a specific form for the relationship between the attributes and the outcome. We would like to simulate a set of attributes from specified marginal distributions and we would like to simulate outcomes from the conditional distribution given the attributes.

%\textcolor{red}{This is a simple matter provided that we are able to specify the parameters of the model that are to be used. Unfortunately, practitioners cannot reasonably be expected to specify the parameters of the model. However, they can generally be relied upon to specify how much more likely default is given a specific attribute than given some other attribute. That is, they can be called upon to specify bad ratios. As a result, we would like to use specified bad ratios to construct a model.}

%\textcolor{red}{Given the emphasis on the interpretability of these models and the requirement that the models used be explained to a non-technical audience, we propose another method to obtain the model parameters based on specified bad ratios.}

\section{Motivating examples}

This section outlines several examples. We start by considering a simple model and we show that the parameters corresponding to a single specified bad ratio can be calculated explicitly, negating the need for the proposed simulation technique. Thereafter we consider slightly more complicated settings and we demonstrate that, in general, no solution exists for a specified set of bad ratios. We also highlight the difficulties encountered when attempting to find the required parameters, should a solution exist. Finally, we consider a realistic model, similar to what one would use in practice.
%The specified model is used to obtain the numerical results in Section \ref{NumRes}.
%For this model, it is not possible to find parameters which match the specifications of the bad ratios exactly.
%\textcolor{red}{as closely as possible. This statement is made precise below.}

It should be noted that we consider both discrete and continuous attributes below. There does not seem to be general consensus between practitioners on whether or not continuous attributes should be included in the model as these attributes are often discretised during the modelling process (some practitioners may argue that we only need consider discrete attributes while others argue against this discretisation); for a discussion, see pages 45 to 56 of \cite{SIDDIQI2006}. Since the number of attributes considered simultaneously using the proposed simulation technique is arbitrary, we may simply chose to replace any continuous attribute by its discretised counterpart. As a result, the techniques described below are applicable in either setting mentioned above.

\subsection*{A simple example}

Let $X_j$ be a single attribute, associated with the $j^{\textrm{th}}$ applicant, with two levels, $0$ and $1$. Denote the respective frequencies with which these values occur by $p$ and $1-p$, respectively;
$$
    X_j=
    \begin{cases}
        1, \textrm{ with probability } p,\\
        0, \textrm{ with probability } 1-p,
    \end{cases}
$$
for $j\in\{1,\dots,n\}$. Let $Y_j$ be the indicator of default for the $j^{\textrm{th}}$ applicant. Denote the overall bad rate by $d$; meaning that the unconditional probability of default is $d:=P(Y=1)$. Let $\gamma$ be the bad ratio of $X_j=1$ relative to $X_j=0$. That is, $\gamma$ is the ratio of the conditional probabilities that $Y_j=1$ given $X_j=1$ and $X_j=0$, respectively; $\gamma:=P(Y_j=1|X_j=1)/P(Y_j=1|X_j=0)$. We may call upon a practitioner to specify appropriate values for $d$ and $\gamma$.

Using the information above, we can calculate the conditional default rates $d_{0}:=P(Y_j=1|X_j=0)$ and $d_{1}:=P(Y_j=1|X_j=1)$. Simple calculations yield
\begin{equation*}
    d_{0} = \frac{d}{p\gamma +1-p}, \ \ \ d_{1}=\frac{d\gamma}{p\gamma +1-p}.
\end{equation*}
In this setting, building a scorecard requires that the following logistic regression model be fitted:
\begin{equation} \label{vgl1}
    \textrm{log}\left(\frac{d_j}{1-d_j}\right) = \beta_0 + j\beta_1, \ \ \ j\in\{0,1\}.
\end{equation}
Calculating the parameters of the model which give rise to the specified bad ratio requires solving the two equations in (\ref{vgl1}) in two unknowns. The required solution is calculated to be
\begin{equation*}
    \beta_0=\textrm{log}\left(\frac{d_0}{1-d_0}\right), \ \ \  \beta_1=\textrm{log}\left(\frac{d_1}{1-d_1}\right)-\beta_0.
\end{equation*}
As a result, given the values of $p$, $d$ and $\gamma$, we can find a model that perfectly mimics the specified overall probability of default as well as the bad ratio. However, the above example is clearly unrealistically simple.

\subsection*{Slightly more complicated settings}

Consider the case where we have three discrete attributes, each with five nominal levels. In this case, the practitioner in question would be required to specify bad ratios for each level of each attribute. This would translate into fifteen equations in fifteen unknowns (since the model would require fifteen parameters in this setting). Solving such a system of equations is already a taxing task, but two points should be emphasised. First, the models used in practice typically have substantially more parameters than fifteen, making the proposition of finding an analytical solution very difficult. Second, there is no guarantee that a solution will exist in this case.

Next, consider the case where a single continuous attribute, say income, is used in the model. When the scorecard is developed, it is common practice to discretise continuous variables such as income into a number of so-called buckets. As a result, the practitioner may suggest, for example, that the population be split into four categories and they may specify a bad ratio for each of these buckets. However, the ``true'' model underlying the data generates income from a continuous distribution and assigns a single parameter to this attribute in the model. Therefore, this example results in a model with a single parameter which needs to be chosen to satisfy four different constraints (in the form of specified bad ratios). Algebraically, this results in an over specified system in which the number of equations exceed the number of unknowns. In general, an over specified system of equations cannot be solved.

The two examples above illustrate that, even in unrealistically simple cases, we may not be able to obtain parameters which result in the specified bad ratios.

\subsection*{A realistic setting}

We now turn our attention to a realistic setting. Consider the case where ten attributes are used; some of which are continuous while others are discrete. For the discrete case, we distinguish between attributes measured on a nominal scale and attributes measured on a ratio scale. An example of an attribute measured on a nominal scale is the application method used by the applicant as the numerical value assigned to this attribute does not allow direct interpretation. On the other hand, the number of credit cards that an applicant has with other credit providers is measured on an ratio scale and the numerical value of this attribute allows direct interpretation. In the model used, we treat discrete attributes measured on a ratio scale in the same way as continuous variables; that is, each of these attributes are associated with a single parameter in the model.

As mentioned above, we consider a model containing ten attributes. However, since several discrete attributes are measured on a nominal scale, the number of parameters in the model exceeds the number of attributes. To be precise, let $l$ denote the number of parameters in the model and let $m$ denote the number of attributes measured. Note that $l \geq m$, with equality holding only if no discrete attributes measured on a nominal scale are present. Let $\boldsymbol{Z}_j=\{Z_{j,1},\dots,Z_{j,l}\}$ be the set of attributes associated with the $j^{\textrm{th}}$ applicant. This vector contains the values of observed continuous and discrete, ratio scaled, attributes. Additionally, $\boldsymbol{Z}_j$ includes dummy variables capturing the information contained in the discrete, nominal scaled, attributes. Define $\pi_j=E[Y_j|\mathbf{Z}_j]$; the conditional probability of default associated with the $j^{\textrm{th}}$ applicant. The model used can be expressed as
\begin{equation}
    \textrm{log}\left(\frac{\pi_j}{1-\pi_j}\right) = \mathbf{Z_j}^{\top}\boldsymbol{\beta}, \label{eqn1}
\end{equation}
where $\boldsymbol{\beta} = (\beta_1,\dots,\beta_l)^\top$ is a vector of $l$ parameters.

The names of the attributes included in the model, as well as the scales on which these attributes are measured can be found in Table \ref{Attributes}. Care has been taken to use attributes which are often included in credit risk scorecards so as to provide a realistic example. For a discussion of the selection of attributes, see pages 60 to 63 of \cite{SIDDIQI2006}. Additionally, Table \ref{Attributes} reports the information value of each attribute; this value measures the ability of a specified attribute to predict the value of the default indicator (higher information values indicate higher levels of predictive ability). Consider a discrete attribute with $k$ levels. Let $D$ be the number of defaults in the data set, let $D_j$ be the number of defaults associated with the $j^{\textrm{th}}$ level of this attribute and let $n_j$ be the total number of observations associated with the $j^{\textrm{th}}$ level of this attribute. In this case, the information value of the variable in question is
\begin{equation*}
    \textrm{IV}=\sum_{j=1}^k \left(\frac{n_j-D_j}{n-D} - \frac{D_j}{D}\right) \textrm{log}\left(\frac{D(n_j-D_j)}{D_j(n-D)}\right).
\end{equation*}
All calculations below are performed in the statistical software R; see \cite{CRAN}.

%\begin{equation*}
%    \textrm{IV}=\sum \textrm{Distribution Goods}_i - \textrm{Distribution Bads}_i * \textrm{log}{\frac{\textrm{Distribution Goods}_i}{\textrm{Distribution Bads}_i} }
%\end{equation*}
%\textcolor{blue}{where the Distribution Goods is the proportion of observations in a bucket associated with a good outcome and the Distribution Bad is the proportion of observations in a bucket associated with a bad outcome. }
%Higher values of IV is preferable, however attributes with an IV of higher than 0.5 should be handled with care.

\begin{table}[!htbp!]%
\caption{The name, measurement scale and information value of the attributes included in the model}
\label{Attributes}
\centering
\small
\begin{tabular}{lll}
\hline
Name & Scale & Information value \\
\hline
Gender & Ordinal & 0.499\\
Existing customer & Ordinal & 0.441\\
Number of enquiries & Ratio  & 0.394\\
Credit cards with other providers & Ratio & 0.515\\
Province of residence & Ordinal & 0.284\\
Application method & Ordinal & 0.222\\
Age & Ratio & 0.164\\
Total amount outstanding & Ratio & 0.083\\
Income & Ratio & 0.182\\
Balance of recent defaults & Ratio & 0.192\\
\hline
\end{tabular}
\end{table}

For the sake of brevity, we only discuss four of the attributes in detail in the main text of the paper. However, the details of the remaining six attributes, including the numerical results obtained, can be found in Appendix A.

We specify the distribution of the attributes below. For each attribute, we also specify the levels used as well as the bad ratio associated with each of these levels. Care has been taken to use realistic distributions and bad ratios in this example. Admittedly, the process of specifying bad rates is subjective, but we base these values on many years of practical experience in credit scoring and we believe that most risk practitioners will consider the chosen values plausible. However, it should be stressed that the modeller is not bound to the specific example used here; the proposed technique is general and the number and distributions of attributes are easily changed. The attributes are treated separately below.

\subsection{Existing customer}

Existing customers are usually assumed to be associated with lower levels of risk than is the case for applicants who are not existing customers. This can be due to the fact that existing customers have already showed their ability to repay credit extended to them in the past, or are more likely to pay the company where they have other products. We specify that $80\%$ of applicants are exiting customers and that the bad ratio is $2.7$ meaning that the probability of default for a new customer is, on average, 2.7 times higher than the probability of default of an existing customer with the same remaining attributes.

%\textcolor{red}{Moet hierdie stukkie in of uit: Whilst outside of the scope of this work, it should be noted that this attribute is often used as a segmentation tool to enable the building of separate models for each level of the attribute. JdP: Ons kan dit maar uithaal. Dis nie verkeerd nie, maar ook nie regtig so relevant nie.}

\subsection{Credit cards with other providers}

This attribute is an indication of the clients exposure to potential credit. A client could, for example, have a low outstanding balance, but through multiple credit cards have access to a large amount of credit. Depending on the type of product being assessed, this could signal higher risk.
%as the client could go into default once they start using the credit cards to pay off existing debt.
Table \ref{CCOth_param1} shows the assumed distribution of this attribute together with the specified bad ratios.

\begin{table}[!htbp!]%
\caption{Credit cards with other providers}
\label{CCOth_param1}
\centering
\small
\begin{tabular}{llcc}
\hline
Group & Description & Proportion & Bad ratio \\
\hline
0 & No credit cards at another provider & 50\%  & 1.0\\
1 & Credit card at another provider & 30\%  & 1.2\\
2 & Credit cards at another provider & 15\% & 1.7\\
3 & Three or more credit cards at another provider & 5\% & 2.5\\
\hline
\end{tabular}
\end{table}  

\subsection{Application method}

The method of application is often found to be a very predictive indicator in credit scorecards. A customer actively seeking credit, especially in the unsecured credit space, is often found to be of a higher risk than customers opting in for credit through an outbound method like a marketing call. We distinguish four different application methods:
\begin{itemize}
    \item Branch - Applications done in the branch.
    \item Online - Application done through an online application channel.
    \item Phone - Applications done through a non-direct channel.
    \item Marketing call - Application done after prompted by the credit provider.
\end{itemize}
Table \ref{AppMeth_param} specifies the distribution of this attribute as well as the associated bad ratios.

\begin{table}[!htbp!]%
\caption{Application method}
\label{AppMeth_param}
\centering
\small
\begin{tabular}{llcc}
\hline
Group & Description & Proportion & Bad ratio \\
\hline
0 & Branch & 30\%  & 1.0\\
1 & Online & 40\%  & 0.5\\
2 & Phone & 15\% & 1.5\\
3 & Marketing Call & 15\% & 0.4\\
\hline
\end{tabular}
\end{table}  

\subsection{Age}

Younger applicants tend to be higher risk, with risk decreasing as the applicants become older. We assume that the ages of applicants are uniformly distributed between 18 and 75 years. We divide these ages into seven groups, see Table \ref{Age_param}.
\begin{table}[!htbp!]%
\caption{Age}
\label{Age_param}
\centering
\small
\begin{tabular}{lcc}
\hline
Group & Proportion & Bad ratio \\
\hline
18 - 21 & 5\% & 1.00\\
21 - 25 & 7\% & 0.85\\
25 - 30 & 9\% & 0.78\\
30 - 45 & 26\% & 0.66\\
45 - 57 & 21\% & 0.50\\
57 - 63 & 11\% & 0.43\\
63 - 75 & 21\% & 0.31\\
\hline
\end{tabular}
\end{table}

As was mentioned above, the remaining attributes are discussed in Appendix A. In the next section, we turn our attention to the proposed simulation technique.

\section{Proposed simulation technique}\label{section3}

Having described the details of the attributes included in the model, we turn our attention to finding a model which results in bad ratios approximately equal to those specified. This is done by simulating a large data set, containing attributes as well as default indicators. Thereafter, the parameters of the scorecard are estimated by fitting a logistic regression model to the simulated data. We demonstrate in Section 4 that the resulting parameters constitute a model which closely corresponds to the specified bad ratios and other characteristics. The steps used to arrive at the parameters for the model as well as, ultimately, a simulated data set are as follows:
\begin{enumerate}
\item Specify the global parameters.
\item Simulate each attribute separately.
\item Combine the simulated attributes.
\item Fit a logistic regression model.
\item Simulate the final default indicators.
\end{enumerate}

It should be noted that the procedure detailed below assumes independence between the attributes. We opt to incorporate this assumption because it is often made in credit scoring in practice. However, augmenting the procedure below to incorporate dependence between attributes is a simple matter. For example, we can drop the assumption of independent attributes by simulating a group of attributes from a specified copula. Although we do not pursue the use of copulas further below, the reader is referred to \cite{Nel2006} for more details.

\subsection{Specify the global parameters}

We specify a fixed, large sample size. It is important that the initial simulated data set be large even in the case where the final simulated sample may be of more modest size as this will reduce the effect of sample variability. We also specify the overall bad rate. It should be noted that overly small bad rates will tend to decrease the information value of the attributes included in the model (for fixed sets of bad ratios). This is due to the difficulty associated with predicting extremely rare events. We use a sample size of 50 000 and an overall bad rate of 10\% to obtain the numerical results shown in the next section.

%The creation of the global parameters is the first step and will impact on the creation of the data set discussed here. The first parameter to select is the population size. As will be seen in the numerical results, a large population size tends to give a more stable outcome.

%The population bad rate needs to be specified to reflect an expected outcome in the scenario for which the data will be used. Where the scenario is not clearly defined, the user can select any bad rate that they would deem appropriate. Care should be given, in this case, to not make the bad rate too high or too low as this could yield unwanted results. For the purpose of PD modelling, a bad rate of between 5\% and 10\% is advisable.

\subsection{Simulate each attribute separately}

The next step entails specifying the marginal distribution as well as the bad ratio associated with each attribute. In the case of discrete attributes, a bad ratio is specified for each of the levels of the attribute. In the case of continuous attributes, the attributes are required to be discretised and a bad ratio is specified for each level of the resulting discrete attribute. Given the marginal distribution and the bad ratios of an attribute, we explicitly calculate the bad rate for each level of the attribute. Consider an attribute with $k$ levels and let $\delta_j$ be the average bad rate associated with the $j^{\textrm{th}}$ level of the attribute for $j \in \{1,\dots,k\}$. In this case
%Each attribute that needs to be created will require this step to be performed. The inputs required for the variable is the following:
%\begin{itemize}
%\item Number of groups
%\item Required distribution for each group
%\item Bad ratio for each group
%\end{itemize}
%First, the implied bad rate for each group is calculated. This bad rate is calculated such that, given the distribution of the attribute over the group, the bad rate will yield an average as specified for the overall bad rate of the population. For each group, i, the bad rate is calculated as:
\[
    \delta_j=\frac{\mu_j d}{\sum_{l=1}^k{\mu_l} p_l}, \ \textrm{where}  \ \ \mu_j = \frac{\gamma_j p_j}{\sum_{l=1}^k{\gamma_l}}.
\]
%\[
%    Bad Rate_i=\frac{BadTemp_i*Total Bad Rate}{\sum_{i=1}^k{BadTemp_i}*Distribution_i}
%\]
%where
%Total Bad Rate is the desired Bad Rate for the portfolio
%\(Distribution_i\) = The distribution for group i
%\[
%    BadTemp_i = \frac{BadRatio_i*Distribution_i}{\sum_{i=1}^k{BadRatio_i}}
%\]
We now simulate a sample of attributes from the specified marginal distribution. Given the values of these attributes, we simulate default indicators from the conditional distribution of these indicators. That is, given that the $j^{\textrm{th}}$ level of the specific attribute is observed, simulate a $1$ for the default indicator with probability $\delta_j$.
%We now simulate values of the attribute under consideration from its marginal distribution. We also simulate a default indicator associated with this attribute; if the level of the attribute is $j$, then this indicator is $1$ with probability $\delta_j$.
%For each observation, a uniform random variable is simulated. The simulated observation is now assigned to the relevant group and the uniform variable simulated for this observation is used to determine whether the outcome variable is observed as good or bad as follows:
%\(Outcome_i= if Uniform_1 < Bad Rate for group associated with i then 1 else 0\)

\subsection{Combine the simulated attributes}

Upon completion of the previous step, we have a realised sample for each of the attributes with a corresponding default indicator. Denoting the sample size by $n$, the expected number of defaults for each attribute is $nd$. However, due to sample variation, the number of defaults simulated for the various attributes will differ, which complicates the process of combining the attributes to form a set of simulated attributes for a (simulated) applicant. In order to overcome this problem, we need to ensure that the number of defaults per attribute are equal.

For each attribute, the number of defaults follows a binomial distribution with parameters $n$ and $d$. As a result, the number of defaults have expected value $nd$ and variance $nd(1-d)$. Therefore, for large values of $n$, the ratio of the expected and simulated number of defaults converges to $1$ in probability. To illustrate the effect of sample variation, consider the following example. If a sample size of $n=10^6$ is used and the overall default rate is set to $5\%$, then the expected number of defaults is $50\ 000$ for each attribute. Due to sample variation, the number of defaults will vary. However, this variation is small when compared to the expected number of defaults; in fact, a $95\%$ confidence interval for the number of defaults is given by $[49\ 572;50\ 428]$. Stated differently, the probability that the simulated number of defaults will be within $1\%$ of the expected number is approximately $97.8\%$ in this case, while the probability that the realised number of defaults differ from the expected number by more than $2\%$ is less than 1 in 200 000.

The examples above indicate that the simulated number of defaults will generally be close to $nd$, and we may assume that changing the simulated number of defaults to exactly $nd$ will not have a large effect on the relationships between the values of the attribute and the default indicator. As a result, we proceed as follows. If the number of defaults exceed $nd$, we arbitrarily replace $1$'s by $0$'s in the default indicator in order to reduce the simulated number of defaults to $nd$. Similarly, if the number of defaults is less than $nd$, we replace $0$'s by $1$'s.

Following the previous step, the number of defaults per attribute are equal and we simply combine these attributes according to the default indicator. That is, in order to arrive at the details of a simulated applicant who defaults, we arbitrarily choose one realisation of each attributed that resulted in default. The same procedure is used to combine the attributes of applicants who do not default.

%Each attribute simulated in the previous step is sorted on the outcome variable. Great care needs to be taken to ensure that the sorting is only done on the outcome variable and not on the attribute itself. The final outcome should be an unsorted column vector where all the observations associated with an outcome of 1 is grouped together and the observations associated with an outcome of 0 is grouped together.

%Once this is done for all observations simulated in the previous step, a single data set is constructed by placing the column vectors next to each other to form a matrix. A new outcome column is created by assigning a value of 1 to the first \(n * BadRate\) observation and a 0 to the rest of the observations.

\subsection{Fit a logistic regression model}

We now have a (large) data set containing all of the required attributes as well as the simulated default indicators. We fit a logistic regression model to this data in order to find a parameter set that mimics the specified bad ratios. That is, we estimate the set of regression coefficients in (\ref{eqn1}). The required estimation is standard and the majority of statistical analysis packages includes a function to perform the required estimation; the results shown below are obtained using the \emph{glm} function in the \emph{Stats} package of R.

%\textcolor{blue}{The logistic regression model is fit using the glm procedure from the Stats package in R.}

\subsection{Simulate the final default indicators}

When considering the data set constructed up to this point, the simulated values for the individual attributes are realised from the marginal distribution specified for that attribute. As a result, we need only concern ourselves with the distribution of the default indicator. We now replace the initial default indicator by an indicator simulated from the conditional distribution given the attributes (which is a simple matter since the required parameter estimates are now available). The simulated values of the attributes together with this default indicator constitute the final data set.

The following link contains R code used for the simulation of a data set using the proposed method; https://bit.ly/3HIMaMU. We emphasise that the user is not bound by the specifications chosen in this paper as the code is easily amended in order to change the distributions of attributes, to specify other bad ratios and to add or remove attributes from the data set.

%estimated parameter values as well as the attributes simulated in step ???, we can calculate the probability of default for each simulated applicant. In the final step, we simulate default indicators which are assigned a value of 1 with the probability of default as specified by the model.

\section{Performance of the fitted model} \label{NumRes}

%\textcolor{red}{Steekproefgrootte 50 000}

%\textcolor{red}{10 000 simulasies.}

%General comments: I am having a little trouble in figuring out the differences between the last two classes of attributes. I think we should consider combining this into a single section? The difficulty comes in when we try to classify other attributes; I think we will be able to find attributes which will fit equally well into both categories.

%Okay... big comment... I do not think we should bother with simulating continuous attributes! These attributes are going to be divided into discrete groups anyway, so won't it be simpler to just simulate from the corresponding discrete proportions anyway? Okay, so an argument can be made to say that we want to change, for instance, the lognormal distribution that is generating the actual income or whatever, but essentially we can then simply calculate the new proportions (outside of the simulation program) and simulate from the corresponding discrete distribution. But I'm sure this recommendation will result in a healthy debate! Either way, this will not change the descriptions provided above, we will simply comment on this after the variables have been described! Just think on this so long if you have a minute please! :-)

In order to illustrate the techniques advocated for above, we use the proposed technique to simulate a number of data sets using the specifications in Section 3. Below, we report the means (denoted ``Observed bad rate'') and standard deviations (denoted ``Std dev of obs bad rate'') of the observed bad ratios obtained when generating $10 \ 000$ data sets, each of size $50 \ 000$.

In Tables 5 to 8. we consider each of the four attributes discussed in the previous section in the main text, while the results associated with the remaining attributes are considered in Appendix B. Tables 5 to 8 indicate that the average observed bad ratios are remarkably close to the nominally specified bad ratios. Furthermore, the standard deviations of the observed bad ratios are also shown to be quite small, indicating that the proposed method results in data sets in which the specifications provided in Section 3 are closely adhered to.

\begin{table}[!htbp!]%
\caption{Exisiting Customers}
\label{ExistCust_Oth_param}
\centering
%\smalls
\begin{tabular}{clccc}
Group & Description & Specified bad rate & Observed bad rate & Std dev of obs bad rate\\
\hline
0 & Yes & 7.46\% & 7.48\% & 0.14\% \\
1 & No & 20.15\% & 20.09\% & 0.46\% \\
\hline
\end{tabular}
\end{table}

\begin{table}[!htbp!]%
\caption{Credit cards with other providers}
\label{CCOth_param}
\centering
%\smalls
\begin{tabular}{clccc}
Group & Description & Specified bad rate & Observed bad rate & Std dev of obs bad rate\\
\hline
0 & No Credit Cards & 4.00\% & 4.70\% & 0.16\% \\
1 & One Credit Card & 12.00\% & 10.43\% & 0.24\% \\
2 & Two Credit Cards & 20.00\% & 19.49\% & 0.46\% \\
3 & Three or more Credit Cards & 28.00\% & 31.90\% & 1.01\% \\
\hline
\end{tabular}
\end{table}
 
\begin{table}[!htbp!]%
\caption{Application Method}
\label{AppMeth_param1}
\centering
\small
\begin{tabular}{clccc}
\hline
Group & Description & Specified bad rate & Observed bad rate & Std dev of obs bad rate\\
\hline
0 & Branch & 12.74\% & 12.73\% & 0.32\% \\
1 & Online & 6.37\% & 6.39\% & 0.21\% \\
2 & Phone & 19.11\% & 19.05\% & 0.55\% \\
3 & Marketing Call & 5.10\% & 5.12\% & 0.34\% \\
\hline
\end{tabular}
\end{table}  

\begin{table}[!htbp!]%
\caption{Age}
\label{Age_param1}
\centering
\small
\begin{tabular}{clccc}
\hline
Group & Description & Specified bad rate & Observed bad rate & Std dev of obs bad rate\\
\hline
0 & 18 - 21 & 17.54\% & 16.82\% & 0.77\% \\
1 & 21 - 25 & 14.91\% & 15.19\% & 0.63\% \\
2 & 25 - 30 & 13.68\% & 13.89\% & 0.53\% \\
3 & 30 - 45 & 11.58\% & 11.46\% & 0.27\% \\
4 & 45 - 57 & 8.77\% & 8.66\% & 0.28\% \\
5 & 57 - 63 & 7.54\% & 7.28\% & 0.37\% \\
6 & 63 - 75 & 5.44\% & 5.82\% & 0.26\% \\
\hline
\end{tabular}
\end{table}

The marginal distributions of the attributes are not reported in the tables since the average observed proportions coincide with the specified proportions up to 0.01\% in all cases. This result is not unexpected, when taking the large sample sizes used into account.

\section{Practical application}

%\textcolor{red}{Ons moet hierdie dalk nog iewers insit: Population stability is tested to determine if the distribution of the attributes of the current population is the same as those of the population on which the scorecard was built. The aim of this step is to determine whether or not the business can be confident in continuing to use the model in its current form. As a result, testing population stability necessarily requires that distribution of the attributes be well understood.}

The method described above provides a way to arrive at a parametric model, which can be used for simulation purposes, via specification of bad ratios for each attribute considered. One interesting application of this procedure is to specify a deviation from the distribution of the attributes and default indicator and to simulate a second data set. This deviation may, for instance, be in the form of specifying a change in the marginal distribution associated with one or more attributes. The newly simulated data set can then be analysed in order to gauge the effect of the change to, for example, the overall credit risk of the population.

In practice, a common metric used to measure the level of population stability is, the aptly named, population stability index (PSI). The PSI quantifies the discrepancy between the observed proportions per level of a given attribute in two samples. Typically, the first data set is observed when the scorecard is developed (we refer to this data set as the \emph{base} data set) and the second is a more recent sample (referred to as the \emph{test} data set). The PSI is calculated as follows;
\begin{equation}
    PSI = \sum_{j=1}^{n} (T_j - B_{j})\log \left(\frac{T_j}{B_j}\right),
\end{equation}
where $T_j$ and $B_j$ respectively represent the proportion of the $j^{\textrm{th}}$ level of the attribute in question in the test and base data sets. The following rule-of-thumb for the interpretation of PSI values is suggested in \cite{SIDDIQI2006}; a value of less than 0.1 indicates that the population shows no substantial changes, a PSI between 0.1 and 0.25 indicates a small change and a PSI of more than 0.25 indicates a substantial change. The PSI is closely related to the Kullback-Leibler distance; for details, see \cite{KL1951} as well as \cite{Kul1959}.

By means of a practical example consider the following setup. A single realisation of the base data set is simulated using the marginal distributions and the bad ratios specified in Section 2 and Appendix A. We also simulate a test data set using the same specifications, with only the following changes:
\begin{itemize}
    \item The proportion of existing customers is changed from 80\% to 57\%. The new distribution is chosen such as to have a PSI value that is approximately 0.25.
    \item The distribution for the number of enquiries is changed from (30\%, 25\%, 20\%, 15\%, 5\%, 5\%) to (10\%, 10\%, 20\%, 50\%, 5\%, 5\%).
\end{itemize}
Following these changes, a test data sets is simulated from the distribution specified above and the resulting PSI is calculated for each attribute. This process is repeated 1~000 times in order to arrive at 1~000 PSI values for each attribute.

In addition to considering the magnitude in the change of the distribution of the attributes, we are interested in measuring the change in the overall credit risk of the population. In order to achieve this, it is standard practice to divide the applicants into various so-called risk buckets based on their probability of default as calculated by the scorecard. In the example used here we proceed as follows; at the time when the data for the base data set is collected, the applicants may be segmented into ten risk buckets, each containing 10\% of the applicants. That is, the $10\%,20\%,\dots,90\%$ quantiles of the probabilities of default of the base data set are calculated. Then, given the test data set, we calculate the proportions of applicants for whom the calculated probability of default is between the $10(j-1)\%$ and $10j\%$ quantiles of the base data set, for $j\in\{1,2,\dots,10\}$. These proportions are then compared to those of the base data set (which are clearly 10\% for each risk bucket) in the same way as the proportions associated with the various levels of the attributes are compared. Table \ref{PSI} contains the average and standard deviations of the PSI calculated for each of the attributes as well as for the risk buckets.

%Once the model is created on the base data, the entire data set is re-scored to get an estimate for each customer. A common practice in industry is to then create a number of buckets, say $k$ for the outcome. These buckets can be used for decision making. Consider a bank that has segmented their credit card scorecard into 10 buckets. Following some analysis, the bank knows that it can make a profit by granting credit cards to the applicants falling in the six lowest risk buckets. The bank further applies its strategy by giving customers in the lowest two risk buckets a better product and higher credit limit than customers in the next four risk buckets. The risk buckets can also be used for determining stability in the outcome of the model. The proportion of customers in each bucket for the base population can be compared to the proportion in each bucket for a test data set using the population stability metrics.

%It is common practice to create ten buckets of equal size, i.e. each bucket contains $10\%$ of the population; however, this is not a requirement and the buckets can be created in any way that makes sense to the business.

\begin{table}[!htbp!]%
\caption{Population Stability Index}
\label{PSI}
\centering
\small
\begin{tabular}{lcc}
\hline
Attribute & Average PSI & Standard dev of PSI \\
\hline
Gender & 0.000 & 0.000\\
Existing customer & 0.256 & 0.010\\
Number of enquiries &  0.800 & 0.018\\
Credit cards with other providers & 0.001 & 0.000\\
Province of residence & 0.001 & 0.001\\
Application method & 0.001 & 0.000\\
Age & 0.001 & 0.000\\
Total amount outstanding & 0.001 & 0.001 \\
Income & 0.001 & 0.000\\
Balance of recent defaults & 0.001 & 0.000\\
Risk buckets & 0.093 & 0.006\\
%Gender & 0.0001 & 0.0002\\
%Existing customer & 0.2557 & 0.0102\\
%Number of enquiries &  0.7988 & 0.0178\\
%Credit cards with other providers & 0.0005 & 0.0004\\
%Province of residence & 0.0012 & 0.0005\\
%Application method & 0.0005 & 0.0004\\
%Age & 0.0008 & 0.0004\\
%Total amount outstanding & 0.0011 & 0.0006 \\
%Income & 0.0008 & 0.0004\\
%Balance of recent defaults & 0.0005 & 0.0003\\
%PD Groups & 0.0926 & 0.0061\\
\hline
\end{tabular}
\end{table}   

When considering the results in Table 9, three observations are in order. First, the PSI values calculated for the risk buckets is less than 0.1, indicating that no substantial change in the distribution of the data is observed.
%Given that it is relatively close to 0.1, the seasoned analyst will investigate the underlying attributes to understand what is driving the change.
Second, the PSI values for the attribute ``existing customer'' is, on average, 0.256.
%and a standard deviation of 0.0102.
Based on the average PSI, the analyst would typically conclude that the variable is unstable as the calculated average PSI value exceeds the cut-off of 0.25. However, in 27.5\% of the simulated test data sets, the PSI was calculated to be less than 0.25. This demonstrates that the proposed simulation technique enables us to perform sensitivity analysis in cases where a change in the distribution of the attributes results in PSI values close to the cut-off value of 0.25. Finally, the only attribute for which there is no doubt that a substantial change has occurred is the ``number of enquiries''. The PSI values calculated for this attribute has an average of 0.8 and a standard deviation of 0.018.

\section{Conclusions}

We propose a simulation technique which can be used in order to generate data sets for use with credit scoring and we specifically demonstrated the usefulness of this technique for testing population stability. The proposed technique is based on the simple idea of specifying bad ratios and finding parameters which approximately adhere to the specified bad ratios. Using a realistic example, we demonstrate that the proposed technique is able to mimic the specified bad ratios with a high degree of accuracy.

%We developed a method allowing us to specify a data set we specify bad rate bad ratios and distributions.
%This method allows an approximation to this.
%This allows many applications:
%We can simulate many realisations from a base and test data sets.
%We can fit a model to observed data.
%Business wise wil jy die vermoe he om iets te doen wat nog nie gedoen is nie want jou strategie gaan verander!

%10 attributes were considered. We specifically look at 4. We see that the specified distributions are followed is shap. Die bad ratios is ook baie naby in die groot meerderheid van die gevalle. Selfs waar hulle uit is is hulle naby.

The proposed simulation method enables one to study the properties of population stability metrics in a systematic manner. This allows for the direct comparison of the various measures commonly used in practice in order to identify the strengths and weaknesses of each; research into this topic is currently underway. The proposed method also simplifies the study of newly proposed tests for population stability. Furthermore, another direction for future research is to generalise the proposed simulation technique to the multivariate case; for instance in the context of multinomial regression.

%Reject inference.
%What are the distributions of population stability metrics (both under H0 and HA)???
%Multivariate stuffies. For instance, multinomial regression.

\section{Appendix A}

%\subsection{Binary attributes}
%The attribute contains only two levels. All data records will contain either one of the levels.

Below, we specify the marginal distributions and the specified bad ratios for the characteristics not discussed in detail in the main text of Section 2. Again, we treat each attribute separately.

\subsection{Gender}

We assume that $60\%$ of applicants are female and $40\%$ are male and we specify the bad ratio of males to females to be $3$.

\subsection{Number of enquiries}

The number of enquiries is a measure of the clients appetite for credit. A client with a large credit appetite will apply for a number of loans. The number of enquiries provides a view of both the clients successful and unsuccessful applications. Higher numbers of enquiries are often associated with increased levels of risk. Table \ref{NumEnq_param} specifies the distribution associated with various levels of this attribute.

%In using this attribute the operational procedures should be considered. A client applying for a home loan through a broker will show enquires at all major banks as the broker aims to get the best deal. This is not necessarily a sign of higher credit risk. This attribute is most used by looking only at unsecured type loans where the increase in enquiries is a clear sign of higher risk.
 
\begin{table}[!htbp!]%
\caption{Number of enquiries}
\label{NumEnq_param}
\centering
\small
\begin{tabular}{clcc}
\hline
Group & Description & Proportion & Bad ratio \\
\hline
0 & No enquiries & 30\%  & 1.0\\
1 & One enquiry & 25\%  & 1.3\\
2 & Two enquiries & 20\% & 1.8\\
3 & Three enquiries & 15\% & 1.9\\
4 & Four enquiries & 5\% & 2.1\\
5 & Five or more enquiries & 5\% & 2.7\\
\hline
\end{tabular}
\end{table}   

\subsection{Province of residence}

Some provinces are greater economic hubs which may result in inhabitants with lower levels of credit risk. Table \ref{Prov_param} shows the marginal distribution as well as bad ratios assumed for the 9 provinces of South Africa.

%[General comment: do not use too many short paragraphs containing only one or two sentences each. You can use paragraphs like these, but this should really only be done every once in a while to emphasise that the content of this particular paragraph is very important.]

\begin{table}[!htbp!]%
\caption{Province of residence}
\label{Prov_param}
\centering
\small
\begin{tabular}{cccc}
\hline
Group & Description & Proportion & Bad ratio \\
\hline
0 & Gauteng & 40\%  & 1.0\\
1 & Western Cape & 30\%  & 0.7\\
2 & KwaZulu Natal & 7\% & 1.8\\
3 & Mpumalanga & 5\% & 1.5\\
4 & North West & 5\% & 3.0\\
5 & Limpopo & 4\% & 2.5\\
6 & Eastern Cape & 4\% & 2.0\\
7 & Northern Cape & 3\% & 4.0\\
8 & Free State & 2\% & 1.2\\
\hline
\end{tabular}
\end{table}   

\subsection{Total amount outstanding}

An applicant's total amount outstanding is an indication of the current indebtedness and provides a view of the client's previous commitments. Excessively low or high levels of this variable may be associated with higher levels of risk; i.e., a customer with no outstanding amount could be a result of not being able to obtain credit while very high levels of this attribute may indicate difficulty in paying current commitments. The marginal distribution specified for this attribute is standard lognormal, rescaled by a factor of 10 000. The lognormal distribution is chosen since its shape is reminiscent of the empirical distribution typically observed in practice, while the scaling factor is incorporated in order to ensure that the numbers used are of a realistic magnitude. The resulting proportions and bad ratios can be found in Table 12.

%Double check the code on this one
 \begin{table}[!htbp!]%
\caption{Total amount outstanding}
\label{AmtOut_param}
\centering
\small
\begin{tabular}{clcc}
\hline
Group & Grouping & Proportion & Bad Ratio \\
\hline
0 & 0 - 5 000 & 24.4\% & 1.0\\
1 & 5 000 - 10 000 & 25.6\% & 1.2\\
2 & 10 000 - 25 000 & 32.0\% & 2.0\\
3 & 25 000 - 100 000 & 16.9\% & 2.1\\
4 & more than 100 000 & 1.1\% & 0.8\\
%4 & 100 000 - 500 000 & 1.1\% & 0.8\\
%5 & 500 000 - 1 000 000 & 0.0\% & 0.5\\
%6 & more than 1 000 000 & 0.0\% & 2.5\\
\hline
\end{tabular}
\end{table}  

%\textcolor{red}{The bad ratio in this table is not monotone. Should be comment on this? We Should probably figure out why that is in the first place....}

%\subsection{Continuous range, clustered}
%The attribute spans a continuous range, however certain natural clusters occur. In order to use these variables, they will be grouped into groups such that the bad rate in each group differs as much as possible from the other groups.

\subsection{Income}

Income is a strong indicator of the ability to repay debt and it is often used directly or indirectly in the scoring process. Direct use occurs through inclusion into the scoring model as an attribute, while indirect use can be accomplished through using income as an entry criteria for the application. The distribution used for income is a mixture with several local models. The associated proportions and bad ratios can be found in Table 13.

\begin{table}[!htbp!]%
\caption{Income}
\label{Inc_param}
\centering
\small
\begin{tabular}{clcc}
\hline
Group & Grouping & Proportion & Bad Ratio \\
\hline
0 & 0 - 5 000 & 3.2\% & 3.0\\
1 & 5 000 - 11 000 & 15.6\% & 2.5\\
2 & 11 000 - 20 000 & 20.4\% & 2.0\\
3 & 20 000 - 30 000 & 21.8\% & 1.4\\
4 & 30 000 - 70 000 & 24.0\% & 1.2\\
5 & more than 70 000 & 15.0\% & 1.0\\
\hline
\end{tabular}
\end{table}

\subsubsection{Balance of recent defaults}

Recent defaults are an indication that a customer is no longer able to pay their debts. This attribute specifically speaks to customers that have recently defaulted as all customer without defaults are grouped at zero. Table 14 specifies a distribution in which the majority of applicants have recent defaults with a value of less than 1 000 units, indicating that the majority of applicants have not defaulted recently.

\begin{table}[!htbp!]%
\caption{Balance of recent defaults}
\label{BalRec_param}
\centering
\small
\begin{tabular}{clcc}
\hline
Group & Grouping & Proportion & Bad ratio \\
\hline
0 & 0 - 1 000 & 60.0\% & 1.0\\
1 & 1 000 - 3 000 & 1.1\% & 1.1\\
2 & 3 000 - 5 000 & 2.1\% & 2.0\\
3 & 5 000 - 30 000 & 18.9\% & 2.5\\
4 & 30 000 - 1 000 000 & 18.0\% & 3.0\\
5 & more than 1 000 000 & 0.0\% & 3.3\\
\hline
\end{tabular}
\end{table}

\section{Appendix B}

Tables 15 to 20 report the specified bad rate, the average observed bad rate as well as the standard deviation of this bad rate for each of the attributes not treated in the main text of Section 4.

\begin{table}[!htbp!]%
\caption{Gender}
\label{GenderTable}
\centering
\small
\begin{tabular}{clccc}
\hline
Group & Description & Specified bad rate & Observed bad rate & Std dev of obs bad rate\\
\hline
0 & Female & 5.56\% & 5.58\% & 0.16\% \\
1 & Male & 16.67\% & 16.62\% & 0.27\% \\
\hline
\end{tabular}
\end{table}

\begin{table}[!htbp!]%
\caption{Number of enquiries}
\label{NumEnq_param1}
\centering
\small
\begin{tabular}{clccc}
\hline
Group & Description & Specified bad rate & Observed bad rate & Std dev of obs bad rate\\
\hline
0 & No Enquiries & 6.62\% & 6.97\% & 0.23\% \\
1 & One Enquiry & 8.61\% & 8.62\% & 0.25\% \\
2 & Two Enquiries & 11.92\% & 10.89\% & 0.30\% \\
3 & Three Enquiries & 12.58\% & 12.74\% & 0.39\% \\
4 & Four Enquiries & 13.91\% & 14.96\% & 0.73\% \\
5 & Five or more Enquiries & 17.88\% & 18.28\% & 0.84\% \\
\hline
\end{tabular}
\end{table}

\begin{table}[!htbp!]%
\caption{Province of residence}
\label{Prov_param1}
\centering
\small
\begin{tabular}{clccc}
\hline
Group & Description & Specified bad rate & Observed bad rate & Std dev of obs bad rate\\
\hline
0 & Gauteng & 7.78\% & 7.79\% & 0.22\% \\
1 & Western Cape & 5.45\% & 5.47\% & 0.24\% \\
2 & KwaZulu Natal & 14.01\% & 14.00\% & 0.74\% \\
3 & Mpumalanga & 11.67\% & 11.67\% & 0.83\% \\
4 & North West & 23.35\% & 23.27\% & 1.05\% \\
5 & Limpopo & 19.46\% & 19.40\% & 1.11\% \\
6 & Eastern Cape & 15.56\% & 15.51\% & 1.05\% \\
7 & Northern Cape & 31.13\% & 30.98\% & 1.50\% \\
8 & Free State & 9.34\% & 9.32\% & 1.22\% \\
\hline
\end{tabular}
\end{table}

\begin{table}[!htbp!]%
\caption{Amount outstanding}
\label{AmtOut_param1}
\centering
\small
\begin{tabular}{clccc}
\hline
Group & Description & Specified bad rate & Observed bad rate & Std dev of obs bad rate\\
\hline
0 & 0 - 5 000 & 6.43\% & 8.52\% & 0.26\% \\
1 & 5 000 - 10 000  & 7.72\% & 9.01\% & 0.25\% \\
2 & 10 000 - 25 000 & 12.86\% & 10.78\% & 0.23\% \\
3 & 25 000 - 100 000 & 13.51\% & 11.93\% & 0.36\% \\
4 & more than 100 000 & 5.15\% & 12.17\% & 2.98\% \\
%4 & 100 000 - 500 000 & 5.21\% & 13.42\% & 1.60\% \\
%5 & 500 000 - 1 000 000 & 3.25\% & 39.58\% & 37.33\% \\
\hline
\end{tabular}
\end{table}

\begin{table}[!htbp!]%
\caption{Income}
\label{Inc_param1}
\centering
\small
\begin{tabular}{clccc}
\hline
Group & Description & Specified bad rate & Observed bad rate & Std dev of obs bad rate\\
\hline
0 & 0 - 5 000 & 19.07\% & 14.51\% & 0.95\% \\
1 & 5 000 - 11 000 & 15.89\% & 12.65\% & 0.44\% \\
2 & 11 000 - 20 000 & 12.71\% & 11.66\% & 0.34\% \\
3 & 20 000 - 30 000 & 8.90\% & 10.28\% & 0.29\% \\
4 & 30 000 - 70 000 & 7.63\% & 9.70\% & 0.30\% \\
5 & more than 70 000  & 6.36\% & 4.08\% & 0.44\% \\
\hline
\end{tabular}
\end{table}

%plot(density(variablename))
%hist(variablename)
%\include...

\begin{table}[!htbp!]%
\caption{Balance of recent defaults}
\label{BalRec_param1}
\centering
\small
\begin{tabular}{clcccc}
\hline
Group & Description & Specified bad rate & Observed bad rate & Std dev of obs bad rate\\
\hline
0 & 1 000 - 3 000  & 6.11\% & 8.41\% & 0.18\% \\
1 & 3 000 - 5 000  & 6.72\% & 4.51\% & 0.91\% \\
2 & 5 000 - 30 000 & 12.22\% & 5.78\% & 0.75\% \\
3 & 30 000 - 1 000 000 & 15.28\% & 8.21\% & 0.32\% \\
4 & more than 1 000 000 & 18.33\% & 17.99\% & 0.45\% \\
\hline
\end{tabular}
\end{table}  

%\textcolor{red}{Iets in die lyn van: Even in the case where the attributes are not particularly strong, the results are still wondrous! Dis in kommentaar op die Information values.}

\bibliographystyle{plain}
\bibliography{lit-AC}  %%% Remove comment to use the external .bib file (using bibtex).

\end{document}